\documentclass[12pt]{article}
\usepackage{fullpage,amssymb,amsmath}
\usepackage[dvips]{epsfig}

\def\##1{{\bf #1}}
\def\=#1{\underline{\underline{#1}}}

\def\+#1{\underline{\bf #1}}
\def\*#1{\underline{\underline{\bf #1}}}

\def\r#1{(\ref{#1})}
\def\l#1{\label{#1}}
\def\c#1{\cite{#1}}

\def\le{\left(}
\def\ri{\right)}
\def\les{\left[}
\def\ris{\right]}
\def\lec{\left\{}
\def\ric{\right\}}

\def\.{\mbox{ \tiny{$^\bullet$} }}

\def\epso{\epsilon_{\scriptscriptstyle 0}}

\def\muo{\mu_{\scriptscriptstyle 0}}

\def\ko{k_{\scriptscriptstyle 0}}
\def\co{c_{\scriptscriptstyle 0}}

\def\eps{\epsilon}

\def\ct{\cos\theta}

\begin{document}

\Large
\begin{center}
{\bf Orthogonal--Phase--Velocity Propagation of Electromagnetic Plane Waves}

\vspace{10mm} \large

Tom G. Mackay\footnote{Corresponding Author. Fax: + 44 131
650 6553; e--mail: T.Mackay@ed.ac.uk.}\\
{\em School of Mathematics,
University of Edinburgh, Edinburgh EH9 3JZ, UK}\\
 Akhlesh  Lakhtakia\\
 {\em CATMAS~---~Computational \& Theoretical
Materials Sciences Group\\ Department of Engineering Science and
Mechanics\\ Pennsylvania State University, University Park, PA
16802--6812, USA}

\end{center}

\vspace{4mm}

\normalsize

\begin{abstract} In an isotropic, homogeneous, nondissipative,
dielectric--magnetic medium that is simply moving with respect
to an inertial reference frame, planewave solutions of the Maxwell curl
postulates can be such that the phase velocity and the time--averaged
Poynting vector are mutually orthogonal. Orthogonal--phase--velocity
propagation thus adds to the conventional positive--phase--velocity propagation
and the recently discovered negative--phase--velocity propagation that
is associated with the phenomenon of negative refraction.

\end{abstract}


\section{Introduction}

Until quite recently, it was commonly held that the time--averaged
Poynting vector and the phase velocity of a plane wave
in any homogeneous and isotropic medium are coparallel \c{DJ,Bel};
furthermore, the angle between the two vectors was
commonly thought to be acute in homogeneous
anisotropic mediums \c{BW}. The successful fabrication of what
are called negative--phase--velocity (NPV) materials\footnote{Some
researchers prefer the infelicitous term {\em left--handed materials}.}~---~in
contrast to their conventional counterparts, the positive--phase--velocity
(PPV) materials~---~led
 to a radical re--think: the angle between the two vectors can be
obtuse and even $180^\circ$. Much has been written on NPV materials
during the past five years, and we refer the interested reader to three
pedagogical reviews \c{LMW03,BKV,SAR} and two Focus Issues of electronic journals
\c{Pen,LM05}.

Could there be materials that are neither of the PPV nor of the
NPV types? We have already provided an affirmative answer to this
question: there can be isotropic dielectric--magnetic materials
wherein the phase velocity is infinite at some frequency \c{LM04}.
In addition to the infinite--phase--velocity  materials, we
provide here a second affirmative answer: materials in which the
phase velocity is orthogonal to the time--averaged Poynting
vector. Such materials may be called orthogonal--phase--velocity
(OPV) materials.

The OPV materials emerge here as a gift of the special theory of relativity \c{Chen,Pappas,ChUn}. Relative
to a co--moving observer, these materials are supposed to be homogeneous, isotropic,
and with dielectric and/or magnetic properties. Relative to an inertial observer
that is not co--moving, planewave propagation may occur such that the OPV
condition is satisfied. The OPV condition, however, will never be satisfied
for the co--moving observer.

The plan of this paper is as follows: Section \ref{analysis} is devoted
to the Minkowski constitutive relations, planewave propagation, and the
OPV condition. Section \ref{num} contains numerical results for an OPV material
that is non--dissipative. Conclusions are presented in Section \ref{con}.
In the notation adopted,
  ${\rm Re}\lec Q \ric$ and ${\rm
Im}\lec Q \ric$ represent the real and imaginary parts,
respectively, of a complex--valued $Q$. The complex conjugate is
written as $Q^*$. Vectors are identified by bold typeface and
3$\times$3 dyadics are double underlined; $\hat{\#v}$ is a unit
vector co--directional with $\#v$; the unit dyadic is $\=I$;  and
$\#r$  denotes the spatial coordinate vector.
 The
permittivity and permeability of free space (i.e., vacuum) are
 $\epso$ and $\muo$, respectively;
 $\co=(\epso\muo)^{-1/2}$ is the speed of light in free space;
 $\omega$ is the angular frequency; and $\ko = \omega/\co$.

\section{Analysis}\label{analysis}

\subsection{ Minkowski constitutive relations}

As in a predecessor paper \c{NPV_STR}, two inertial reference
frames, $\Sigma$ and $\Sigma'$, provide the setting for our
analysis. The inertial reference frame $\Sigma'$ moves at constant
velocity $\#v=v\hat{\#v}$ relative to the inertial reference frame
 $\Sigma$. The electromagnetic field phasors in frames
$\Sigma$ and $\Sigma'$ are represented as $\lec \#D, \#E, \#H, \#B
\ric$ and $\lec \#D', \#E', \#H', \#B' \ric$, respectively. The
two sets of phasors are related via the Lorentz transformation as
\c{Chen}
\begin{eqnarray}
\#E' &=& \le \#E \.\hat{\#v} \ri \hat{\#v} +
 \frac{1}{\sqrt{1 - \beta^2}} \,
\les \le \=I - \hat{\#v}\hat{\#v} \ri \. \#E + \#v \times
\#B \ris, \l{Ep}  \\
\#B' &=& \le \#B \.\hat{\#v} \ri \hat{\#v} +
 \frac{1}{\sqrt{1 - \beta^2}} \,
\les \le \=I - \hat{\#v}\hat{\#v} \ri \. \#B - \frac{ \#v \times
\#E}{\co^2} \ris,  \\
\#H' &=& \le \#H \.\hat{\#v} \ri \hat{\#v} +
 \frac{1}{\sqrt{1 - \beta^2}} \,
\les \le \=I - \hat{\#v}\hat{\#v} \ri \. \#H - \#v \times
\#D \ris,  \\
\#D' &=& \le \#D \.\hat{\#v} \ri \hat{\#v} +
 \frac{1}{\sqrt{1 - \beta^2}} \,
\les \le \=I - \hat{\#v}\hat{\#v} \ri \. \#D + \frac{ \#v \times
\#H}{\co^2} \ris,  \l{Dp}
\end{eqnarray}
where $\beta = v / \co$.

We consider a homogeneous medium which is characterized by the frequency--domain
dielectric--magnetic constitutive relations
\begin{equation}
\left.
\begin{array}{l}
\#D' =   \epso \eps_r\#E'  \\ \vspace{-3mm} \\
\#B' =  \muo \mu_r\#B'
\end{array}
\right\}, \l{Con_rel_p}
\end{equation}
in the co--moving reference frame $\Sigma'$, with $\eps_r$ as the
relative permittivity and $\mu_r$ as the relative permeability.
 By virtue of \r{Ep}--\r{Dp},
the constitutive relations \r{Con_rel_p} transform to the
Minkowski constitutive relations
\begin{equation}
\left.
\begin{array}{l}
\displaystyle{ \#D = \epso \eps_r\, \=\alpha\.\#E + \frac{m
\hat{\#v}\times\#H}{\co} }
\\ \vspace{-3mm} \\
\#B = \displaystyle{ -\, \frac{m \hat{\#v}\times\#E}{\co} + \muo
\mu_r\, \=\alpha\.\#H\ }
\end{array}
\label{cr} \right\},
\end{equation}
in the non--co--moving
reference frame $\Sigma$; here,
\begin{eqnarray}
\=\alpha &=& \alpha\,\=I + (1-\alpha)\,\hat{\#v}\hat{\#v} \,,\\
\alpha & = & \frac{1-\beta^2}{1- \eps_r\mu_r\beta^2} \,,\\
m &=& \beta\,\frac{ \eps_r\mu_r-1} {1- \eps_r\mu_r\beta^2}\,.
\end{eqnarray}
The constitutive relations \r{cr} degenerate
to  \r{Con_rel_p} as $v
\rightarrow 0$.

In the remainder of this paper, the propagation of plane waves in
a medium described by the Minkowski constitutive relations \r{cr}
is investigated, with particular emphasis on the unusual behaviour
which occurs when $\eps_r\mu_r< 1$. For further general background
details on planewave propagation within the context of special
relativity, the reader is referred to standard works
\c{Chen}--\c{ChUn}.

\subsection{Planewave propagation}

Our focus lies on plane
waves with field phasors
\begin{equation}
\left.
\begin{array}{l}
\#E = \#E_0 \exp \le i \#k \. \#r \ri \\ \vspace{-3mm} \\
\#H = \#H_0 \exp \le i \#k \. \#r \ri
\end{array}
\right\}.  \l{pw}
\end{equation}
The plane waves are taken to be uniform; i.e., the wavevector $\#k
= k \, \hat{\#k}$ where the unit vector $\hat{\#k} \in
\mathbb{R}^3$ and \begin{equation}
k = k_R + i k_I\,,
\end{equation}
where $k_R = \mbox{Re} \lec k \ric$ and $k_I = \mbox{Im} \lec k
\ric$.

As a result of combining the frequency--domain Maxwell curl postulates
with the Minkowski constitutive relations and substituting \r{pw} therein, we find
that \c{NPV_STR}
\begin{equation}
k= \ko \, \frac{-\beta\xi \, \hat{\#k}\.\hat{\#v}  \pm
\sqrt{\Delta}} {1-\xi\le \beta \, \hat{\#k}\.\hat{\#v} \ri^2}\,,
\l{k_sqrt}
\end{equation}
wherein
\begin{eqnarray}
\xi &=& \frac{ \eps_r\mu_r-1} {1-\beta^2}\,,\\
\Delta &=& 1 + \le \eps_r\mu_r- 1 \ri \delta\,,\\
\delta &=& \frac{1 - \le \beta \, \hat{\#k}\.\hat{\#v} \ri^2 }{1-
\beta^2} \ge 1\,.
\end{eqnarray}
Since the two wavenumbers represented by \r{k_sqrt} are not
independent, the medium with constitutive relations \r{cr} is
unirefringent. The choice of sign for the square root term in
\r{k_sqrt} is dictated by the direction of planewave propagation.

The amplitudes of the  electromagnetic field phasors turn out to be \c{NPV_STR}
\begin{eqnarray}
\#E_0 &=& C_1\,\#e_1 + C_2 \,\#e_2 \,, \l{E_vec}\\
\#H_0 &=&  \frac{C_1}{\omega\muo \mu_r}\,\#e_2 -\omega\epso
\eps_r\,C_2 \,\#e_1 \,, \l{H_vec}
\end{eqnarray}
where the orthogonal eigenvectors $\#e_1$ and $\#e_2$ are given by
\begin{eqnarray}
\#e_1& =& \#k\times\hat{\#v}\,, \l{e1_vector}
\\[5pt]
\#e_2 &=& \=\alpha^{-1}\.\les (\#k +
m\,\frac{\omega}{\co}\,\hat{\#v})\times\#e_1\ris\,, \l{e2_vector}
\end{eqnarray}
whereas $C_1$ and $C_2$ are arbitrary constants.

Exploiting the relation
\begin{equation}
\#e_2 = \#a \times \#e_1\,, \l{e2_P}
\end{equation}
where
\begin{equation}
\#a = \#k + \frac{\xi \le \omega - \#k \. \#v \ri}{\co^2} \,
\#v\,,
\end{equation}
 we find the time--averaged Poynting vector
\begin{equation}
\#P = \frac{ | \#e_1 |^2  \exp \le -2 k_I \hat{\#k}\.\#r \ri}{2}
\, {\rm Re} \lec \frac{| C_1 |^2}{\omega \muo \mu^*_r} \#a^* +
|C_2|^2 \omega \epso \eps^*_r \#a \ric\,. \l{P_general}
\end{equation}

\subsection{Nondissipative mediums}

For nondissipative mediums (i.e., $\eps_r\in \mathbb{R}$ and $
\mu_r\in \mathbb{R}$), the following three scenarios may be
identified:
\begin{itemize}

\item[(i)] If $\eps_r\mu_r\ge 1$ then $\Delta > 0$ and the expression \r{k_sqrt}
represents two real-valued wavenumbers.

\item[(ii)] If $0 < \eps_r\mu_r< 1$ then $\Delta$ can be either
positive-- or negative--valued depending upon the magnitude of
$\beta$ and the relative orientations of $\hat{\#k}$ and
$\hat{\#v}$. Hence, either two real--valued wavenumbers or a
conjugate pair of complex--valued wavenumbers are  represented by
\r{k_sqrt}. In the instance where $v = 0$, we have $\Delta =
\eps_r\mu_r> 0$ and the wavenumbers are accordingly real--valued.

\item[(iii)] If $\eps_r\mu_r< 0$
then $\Delta < 0$ and   \r{k_sqrt} provides two complex--valued
wavenumbers. In the case where $v = 0$, we have $k_R = 0$ which
does not correspond to a propagating wave. We exclude this
scenario from further consideration.
\end{itemize}

\subsection{Positive, negative and orthogonal phase velocity}

Equation \r{P_general} for
the time--averaged Poynting vector  simplifies
to
\begin{equation}
\#P = \frac{ | \#e_1 |^2  \exp \le -2 k_I \hat{\#k}\.\#r \ri}{2}
\le \frac{| C_1 |^2}{\omega \muo \mu_r} + |C_2|^2 \omega \epso
\eps_r\ri \les
  k_R \hat{\#k} + \frac{\xi \le \omega -
k_R \hat{\#k} \. \#v \ri}{\co^2} \#v \ris  \,,\l{P_nondissip}
\end{equation}
when $\eps_r\in \mathbb{R}$, $ \mu_r\in \mathbb{R}$, and
$k\in\mathbb{C}$. The phase velocity
\begin{equation}
\#v_p = \co  \frac{\ko}{k_R} \hat{ \#k} \,, \l{vp_def}
\end{equation}
is described as being
 positive, negative or orthogonal depending upon its
 orientation relative to the  time--averaged
Poynting vector. That is, positive phase velocity (PPV) is
characterized by $\#v_p \. \#P > 0$, negative phase velocity (NPV)
is characterized by $\#v_p \. \#P < 0$, and orthogonal phase
velocity (OPV) is characterized by $\#v_p \. \#P = 0$. In light of
\r{P_nondissip},  the phase velocity for nondissipative mediums
 is positive, negative or  orthogonal
depending upon the sign of the scalar
\begin{equation}
W = k_R \les k_R + \xi \beta \, \le \ko - k_R \beta \,
\hat{\#k}\.\hat{\#v} \ri \hat{\#k}\.\hat{\#v} \ris\,. \l{W_scalar}
\end{equation}

We distinguish between the scenarios $\Delta \ge 0$ and $\Delta <
0$ for nondissipative mediums, as follows.

\begin{itemize}
\item[(a)] $\Delta \ge 0$. Here the wavenumber is real--valued and
\begin{eqnarray}
&& W = k \les k + \xi \beta \, \le \ko - k \beta \ct  \ri   \ct \,
\ris \,, \l{P_components_1}
\end{eqnarray}
with $\theta$ being the angle between $\hat{\#k}$ and $\hat{\#v}$.

\item[(b)] $\Delta <0$. Here the wavenumber is complex--valued,
with real and imaginary parts  given by
\begin{equation}
\left.
\begin{array}{l}
 k_R = \ko \, \displaystyle{ \frac{-\beta\xi \, \ct}
{1-\xi\le \beta \, \ct \, \ri^2}\,} \\ \vspace{-3mm} \\ k_I = \ko
\, \displaystyle{\frac{  \pm \sqrt{\Delta}} {1-\xi\le \beta \, \ct
\, \ri^2}}\,
\end{array}
\right\}
 , \l{k_R_I}
\end{equation}
and $W$ is null--valued.
\end{itemize}

\section{Numerical illustrations}\label{num}

Let us now  illustrate the theory presented in the preceding
section by means of some numerical results. In Figure~1, the real
and imaginary wavenumbers are plotted as functions of the
real--valued product $\eps_r\mu_r$ for $\beta \in \lec 0.3, 0.6,
0.9 \ric$ with $\theta \in \lec \pi/6, \pi/3, 3 \pi/4 \ric$,
wherein $\theta$ denotes the angle between $\hat{\#k}$ and
$\hat{\#v}$. The wavenumber generally has a nonzero imaginary
part, despite $\eps$ and $\mu$ being real--valued. Furthermore,
the positive--valued upper bound  on the range of $\eps_r\mu_r$
values for which $k_I \neq 0$ increases as the relative speed
$\beta$ increases.

The magnitudes of the phase velocities, corresponding to the wavenumbers presented
in Figure~1, are plotted in Figure~2. By definition, the phase
velocity is aligned with the direction of $\hat{\# k }$.

The positive, negative and orthogonal nature of the phase
velocity, as deduced from the sign of the scalar $W$ in
\r{W_scalar}, are mapped  in Figure~2 as functions of $\beta$ and
$\theta$ for
\begin{equation*}
\eps_r\mu_r\in \lec  0.2,  0.4, 0.6,  0.8, 1.0, 1.2, 1.5, 2, 3, 4,
5, 10 \ric\,.
\end{equation*}
The OPV region   occupies a substantial portion of the $\beta
\theta$ plane for small values of $\eps_r\mu_r$. As $\eps_r\mu_r$
increases towards unity, the OPV region decreases and vanishes in
the limit $\eps_r\mu_r\rightarrow 1$. The OPV region is
 symmetric
about the line $\theta = \pi/2$, whereas the NPV region is
restricted to those regions of the $\beta \theta$ plane for which
$\pi/2 < \theta \leq \pi$ \c{NPV_STR}~---~which shows that  NPV arises only when the
velocity vector $\#v$ casts a negative projection onto the
wavevector $\#k$.
 When $\eps_r\mu_r= 1$, the phase velocity
is positive for all values of $\beta$ and $\theta$. For
$\eps_r\mu_r> 1$, OPV does not arise: the phase velocity is either
positive or negative, with the   NPV region expanding as the
product $\eps_r\mu_r$ increases.

\section{Concluding remarks}\label{con}

Consider a homogeneous medium which is an isotropic
dielectric--magnetic medium in a co--moving reference frame
$\Sigma'$, characterized by the  relative permittivity scalar
$\eps_r
> 0$ and the relative permeability scalar $ \mu_r> 0$.
From this perspective, the medium supports planewave propagation
with real--valued wavenumbers, and the phase velocity is directed
parallel to the time--averaged Poynting vector.

The nature of planewave propagation is different when viewed in an
inertial reference frame $\Sigma$, which moves uniformly at
velocity $-\#v$ relative to $\Sigma'$. The following two regimes
are then distinguished:

\begin{itemize}

\item[(a)] If $\eps_r\mu_r\ge 1$ then all wavenumbers are
real--valued, regardless of $\#v$. The projection of the phase
velocity onto the time--averaged Poynting vector can be
either positive or negative, depending upon the magnitude and
direction of $\#v$.

\item[(b)]
 If $0 < \eps_r\mu_r< 1$ then
planewave propagation characterized by both  $k \in \mathbb{R}$
and
 $k \in \mathbb{C}$ (with $k_I \neq 0$) is supported. Whether
$ k_I \neq 0$ or not depends upon the magnitude  of $\#v$ and its
orientation relative to $\#k$.
 For
 plane waves with $ k_I \neq 0$, the time--averaged Poynting vector is
 directed orthogonal to the  phase velocity.

\end{itemize}

The regime $0 < \eps_r\mu_r< 1$ lies between either vacuum
($\eps_r=\mu_r=1$) or anti--vacuum ($\eps_r=\mu_r=-1$) and
nihility ($\eps_r=\mu_r=0$) \c{Ltrin,Ziol,LMtrin}. Anti--vacuum is
the vaunted material for the so--called perfect lenses \c{Pen04}.
Many schemes for materials satisfying the condition $0 <
\eps_r\mu_r< 1$ have been formulated \c{Rachford,SH03,PO03},
although we must note that dissipation then is hard to avoid.
However, our calculations (not reported here) show that the OPV
condition can be very nearly satisfied by weakly dissipative
materials.

\newpage

\begin{figure}[!ht]
\centering \psfull \epsfig{file=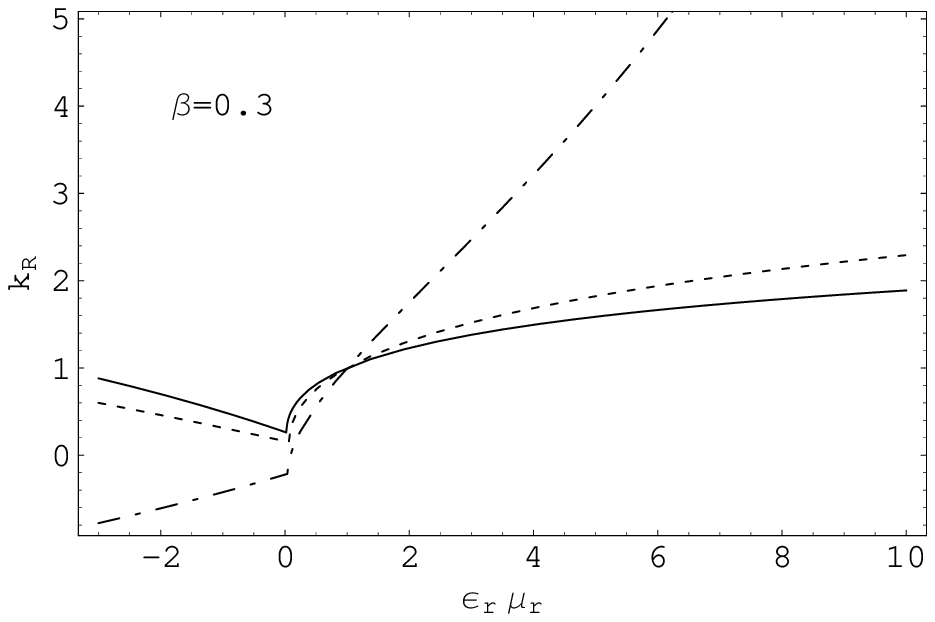,width=3.0in} \hfill
\epsfig{file=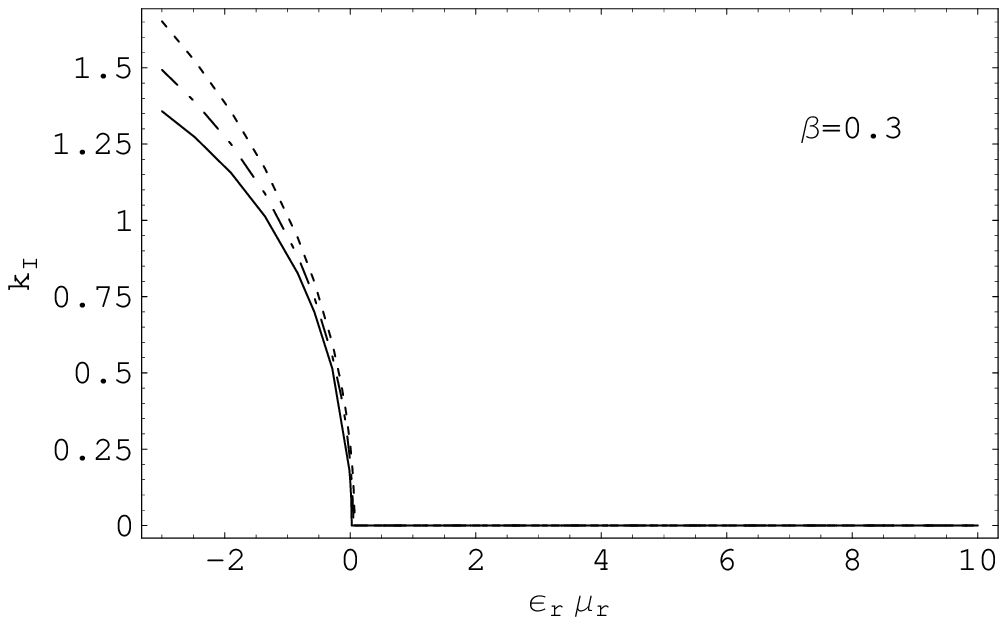,width=3.0in}
\\ \vspace{5mm}
\epsfig{file=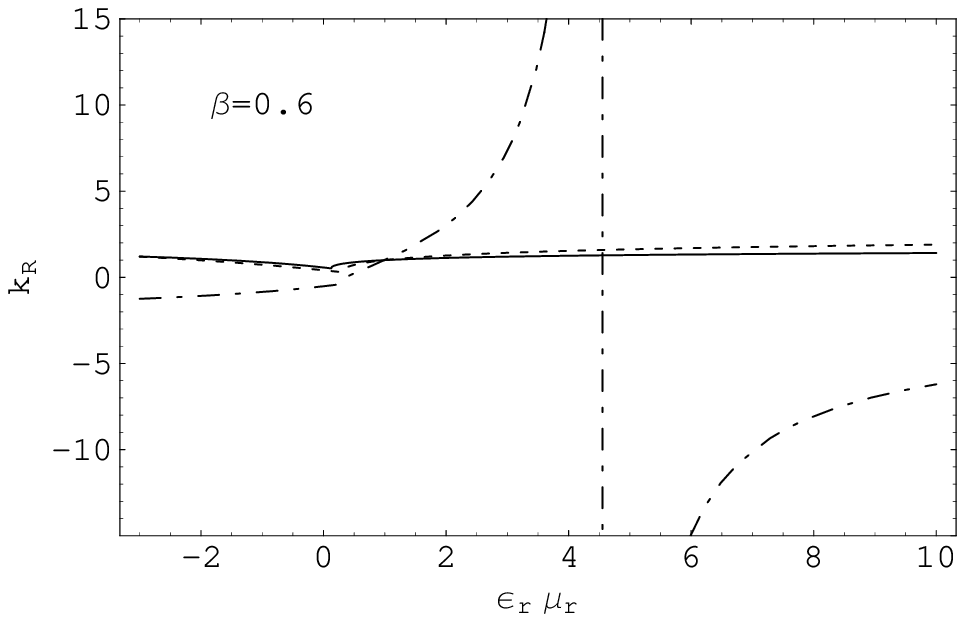,width=3.0in} \hfill
\epsfig{file=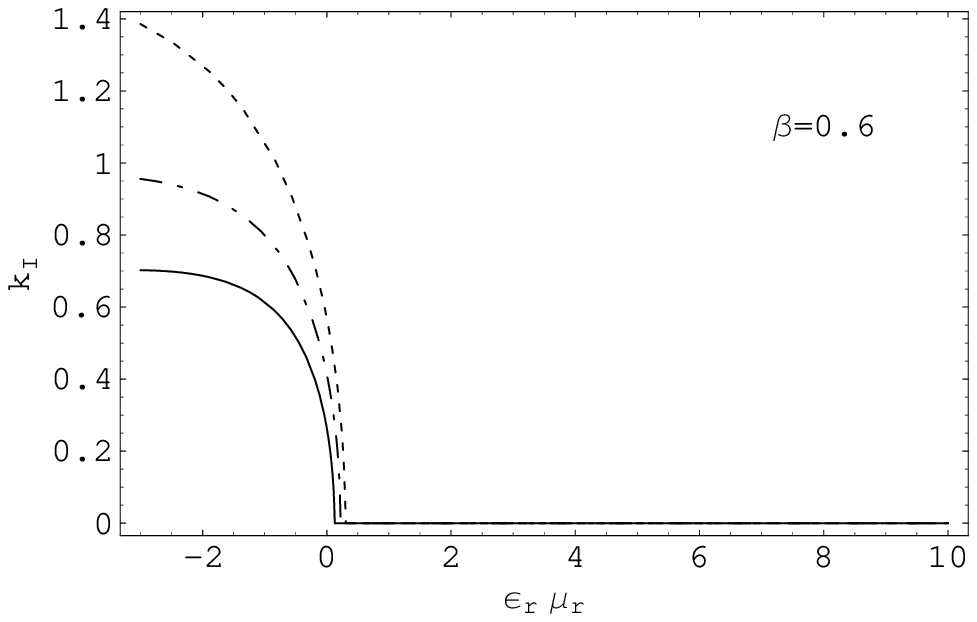,width=3.0in}
\\ \vspace{5mm}
\epsfig{file=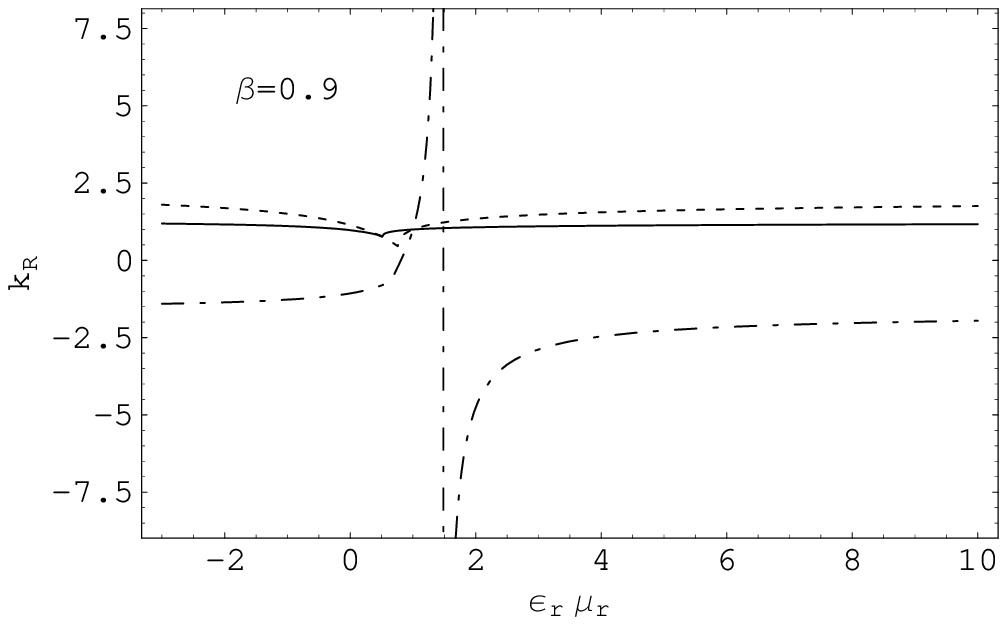,width=3.0in} \hfill
\epsfig{file=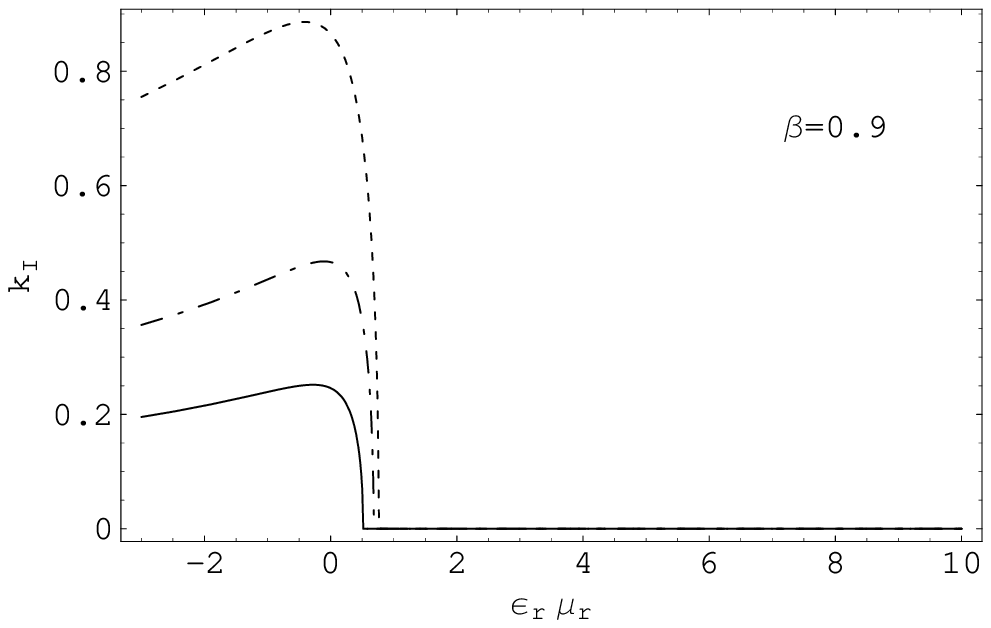,width=3.0in}
  \caption{\label{fig1}
The real (left) and imaginary (right) parts of the  wavenumbers
$k$, normalized with respect to $ \ko$, plotted against
$\epsilon_r \mu_r$  for $\beta \in \lec 0.3, 0.6, 0.9 \ric$. The
values of $\theta$ are: $30^\circ$ (solid curves), $60^\circ$
(dashed curves) and $135^\circ$ (broken dashed curves).
 }
\end{figure}

\newpage

\begin{figure}[!ht]
\centering \psfull \epsfig{file=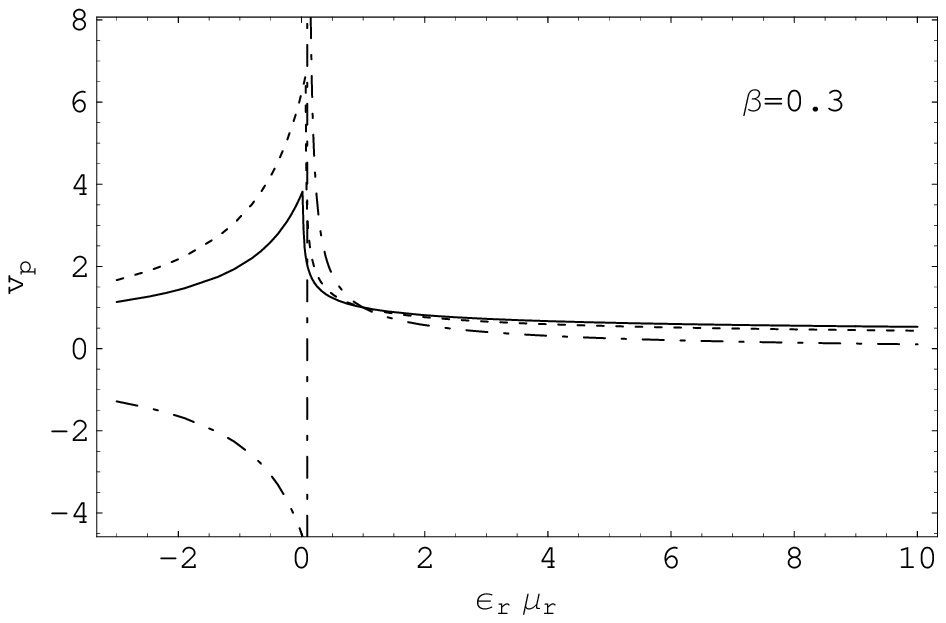,width=4.0in}
\\ \vspace{5mm}
\epsfig{file=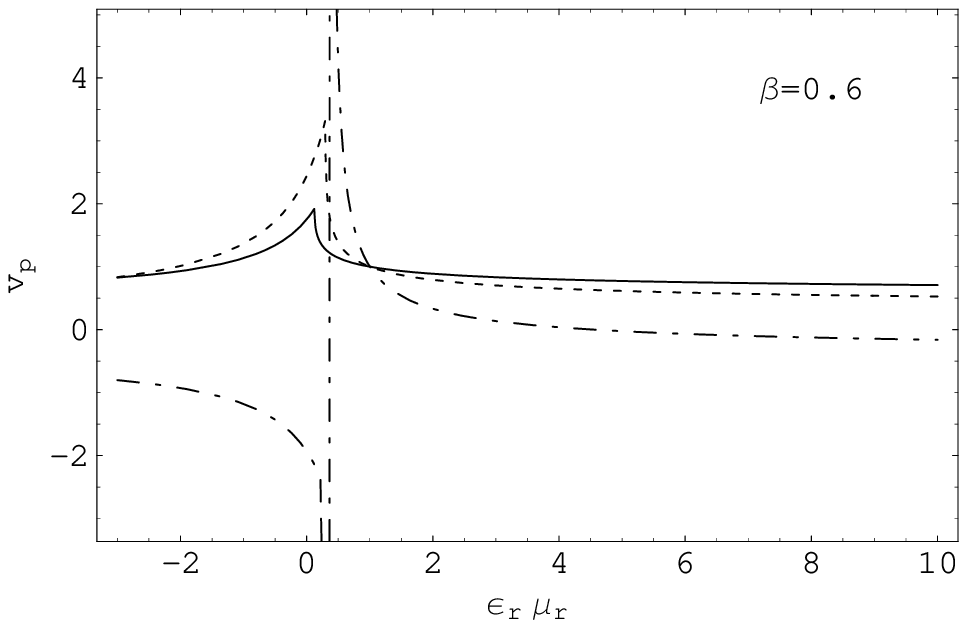,width=4.0in}
\\ \vspace{5mm}
\epsfig{file=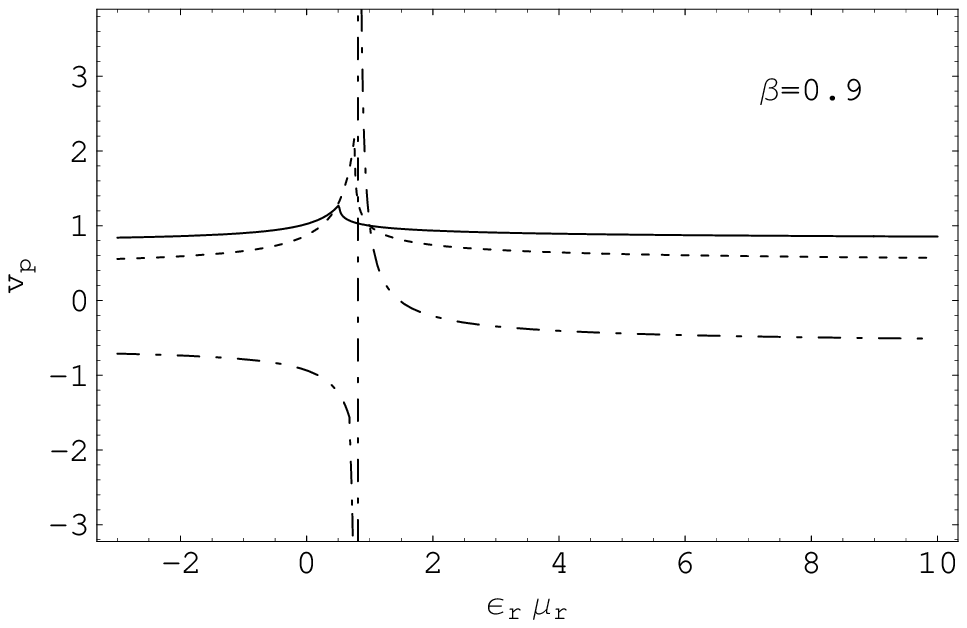,width=4.0in}
  \caption{\label{fig2}
The phase velocity, normalized with respect to $ \co$, plotted
against $\epsilon_r \mu_r$
 for $\beta
\in \lec 0.3, 0.6, 0.9 \ric$.  The values of $\theta$ are:
$30^\circ$ (solid curves), $60^\circ$ (dashed curves) and
$135^\circ$ (broken dashed curves). The phase velocity is directed
along $\hat{{\bf k}}$.
 }
\end{figure}

\newpage

\begin{figure}[!ht]
\centering \psfull \epsfig{file=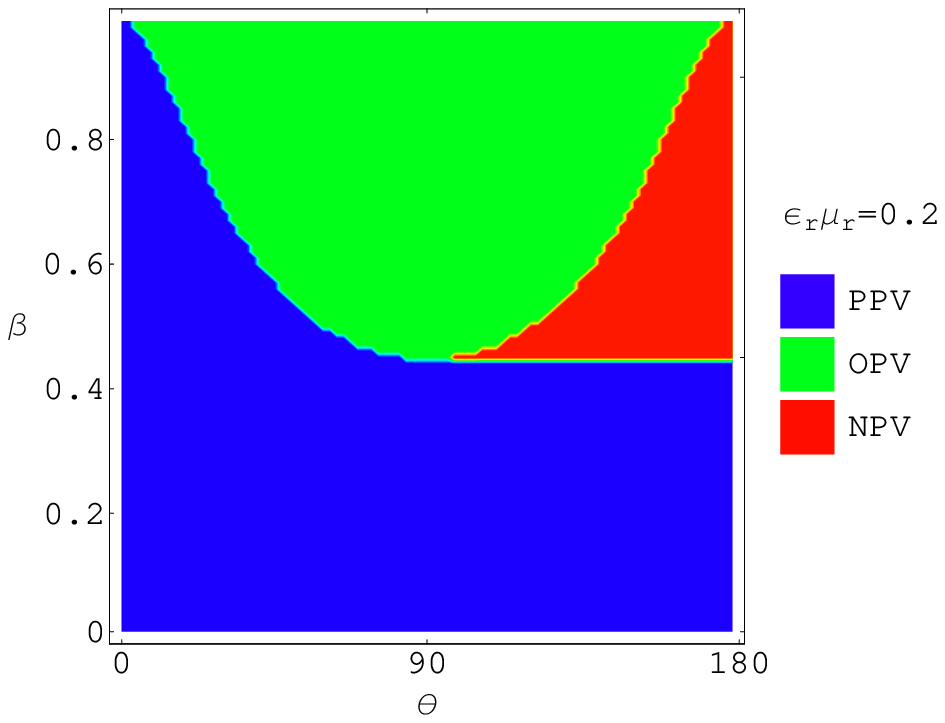,width=2.1in}
\epsfig{file=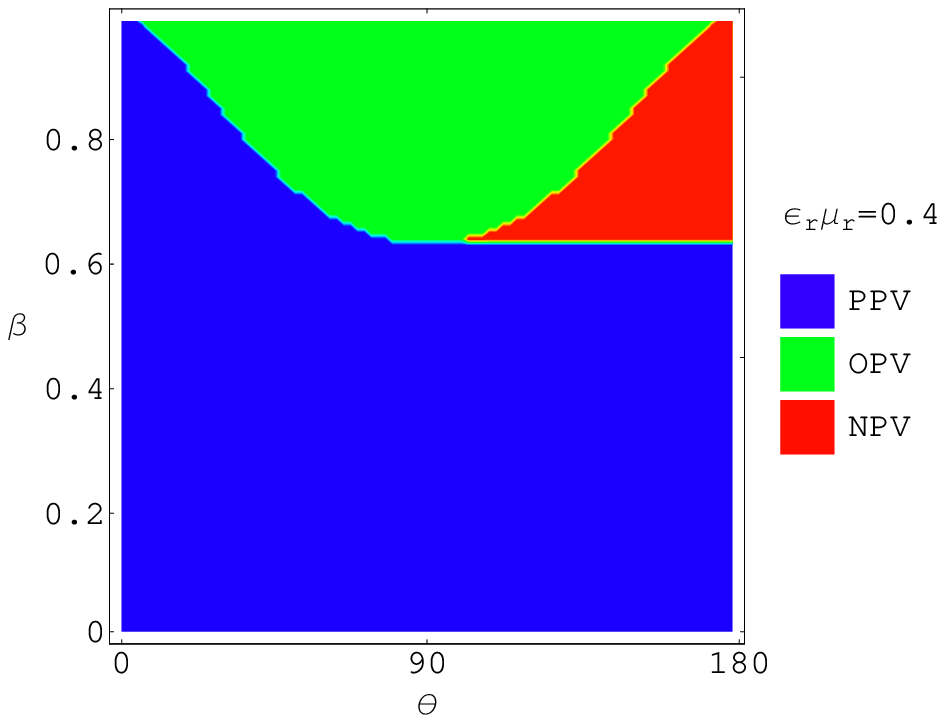,width=2.1in}
\epsfig{file=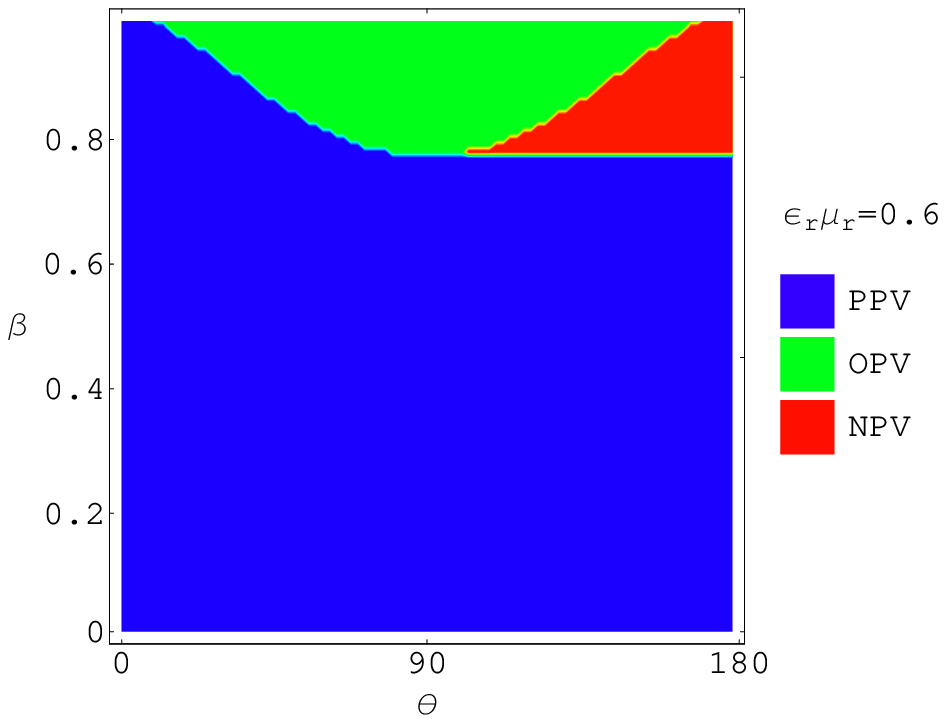,width=2.1in}
\\ \vspace{5mm}
\epsfig{file=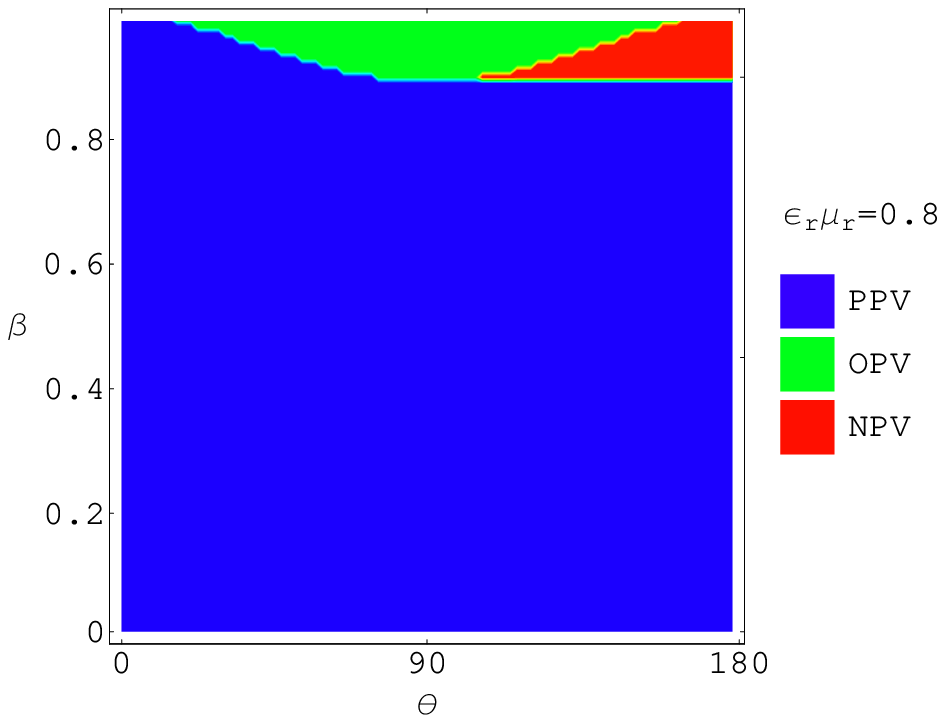,width=2.1in}
\epsfig{file=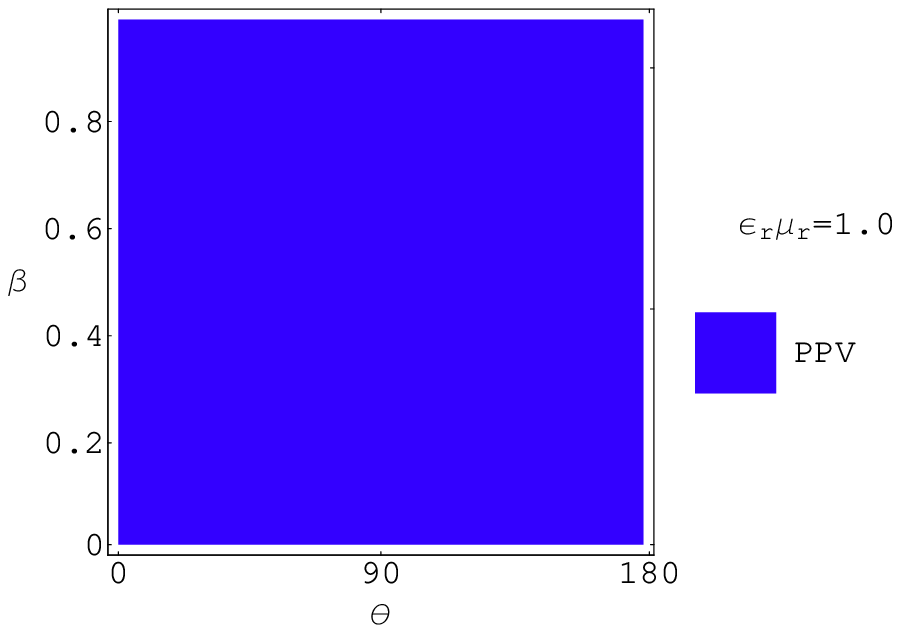,width=2.1in}
\epsfig{file=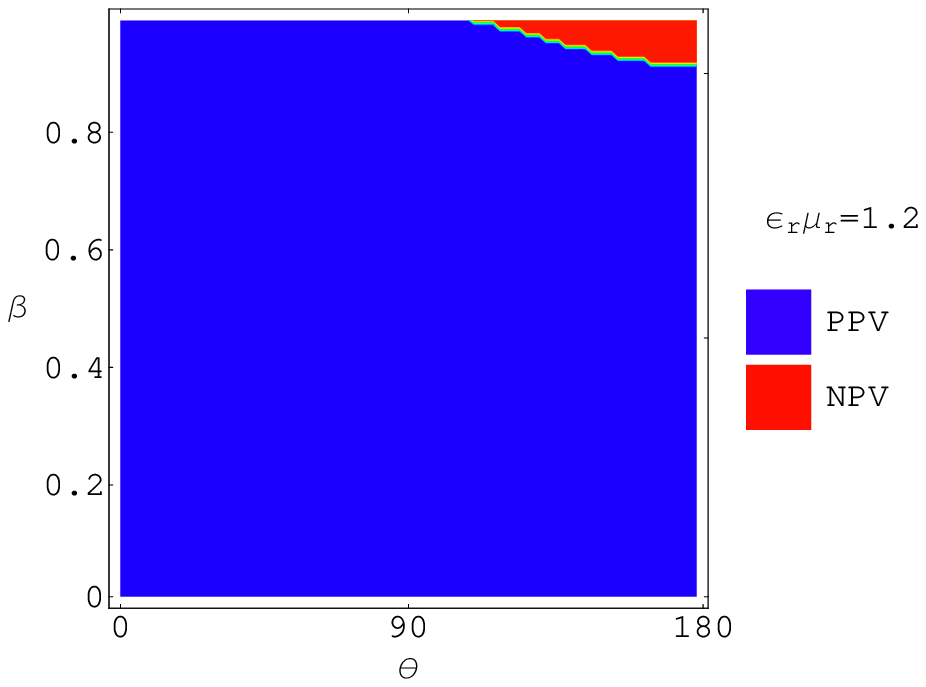,width=2.1in}
  \caption{\label{fig3}
  The distribution of positive phase velocity (PPV), negative
  phase velocity (NPV) and orthogonal phase velocity (OPV),
   as functions of $\beta \in (0,1)$ and $\theta \in (0^\circ,
180^\circ)$
 for
 $
\epsilon_r \mu_r= 0.2$, 0.4, 0.6 , 0.8, 1, 1.2, 1.5, 2, 3, 4, 5,
10. The labels `NPV' and `PPV' shown hold for $\eps_r>0$ and $
\mu_r> 0$, but must be
 reversed for
$\eps_r<0$ and $ \mu_r< 0$.
 }
\end{figure}

\newpage
\setcounter{figure}{2}
\begin{figure}[!ht]
\centering \psfull \epsfig{file=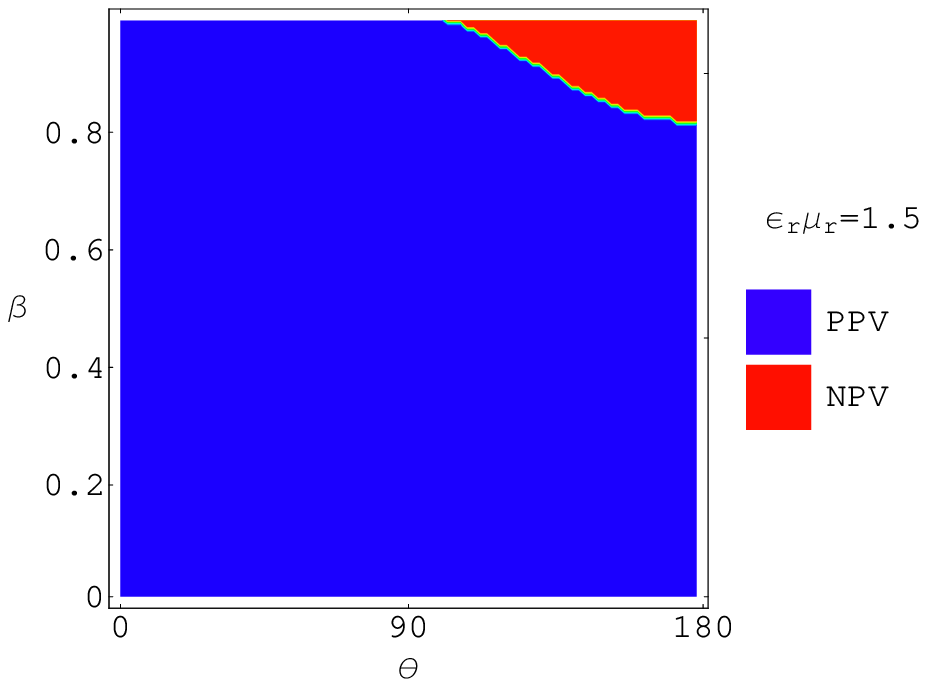,width=2.1in}
\epsfig{file=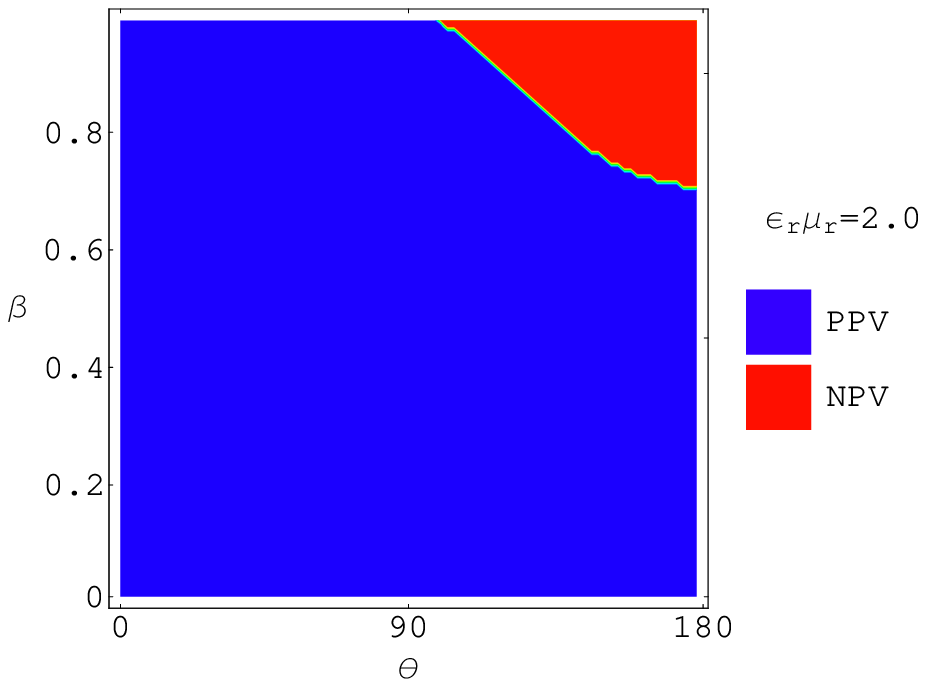,width=2.1in}
\epsfig{file=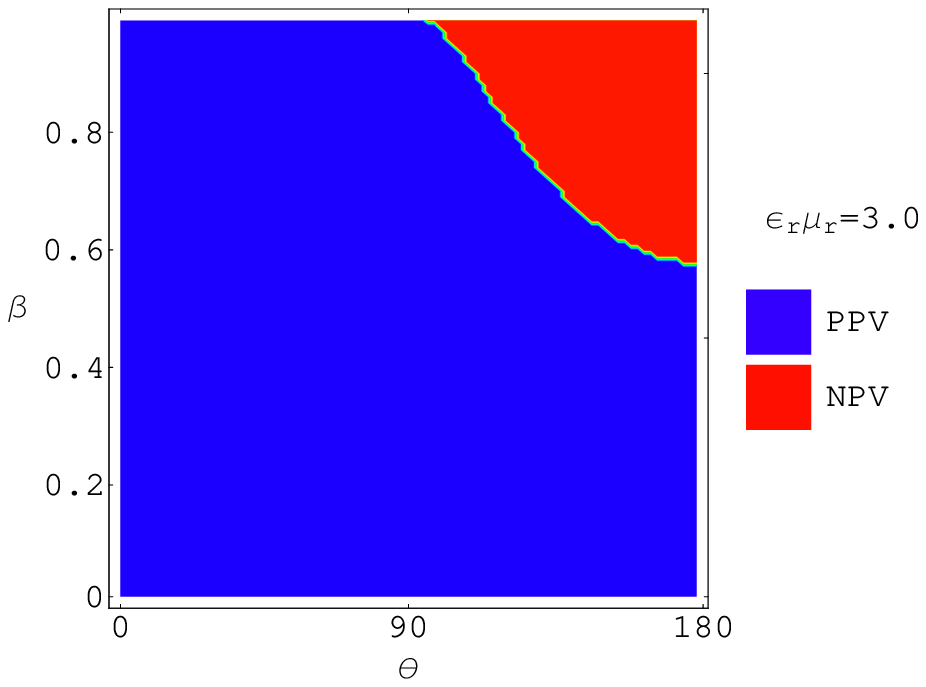,width=2.1in}
\\ \vspace{5mm}
\epsfig{file=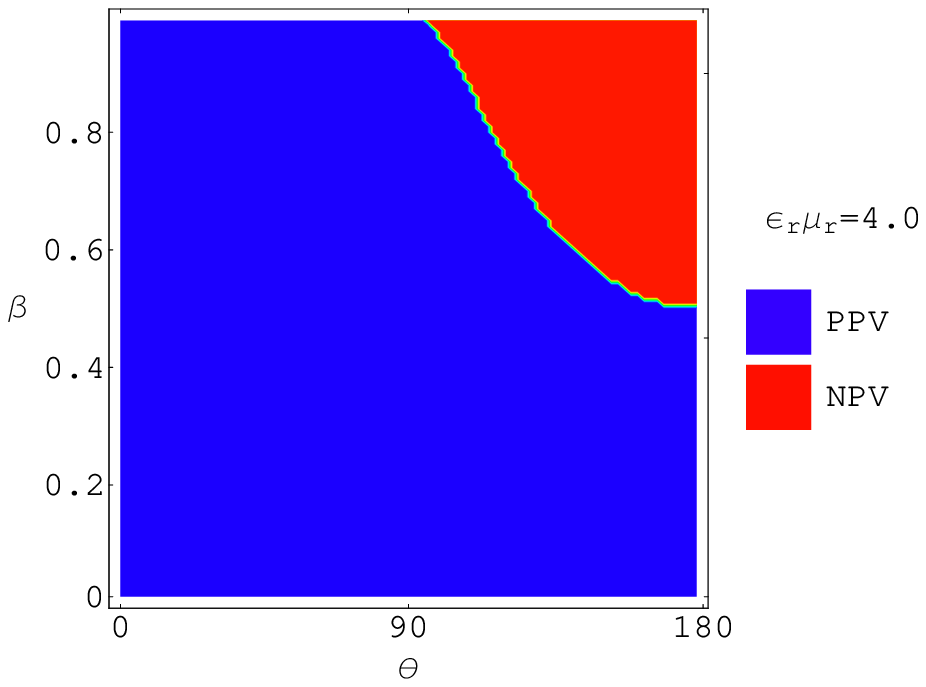,width=2.1in}
\epsfig{file=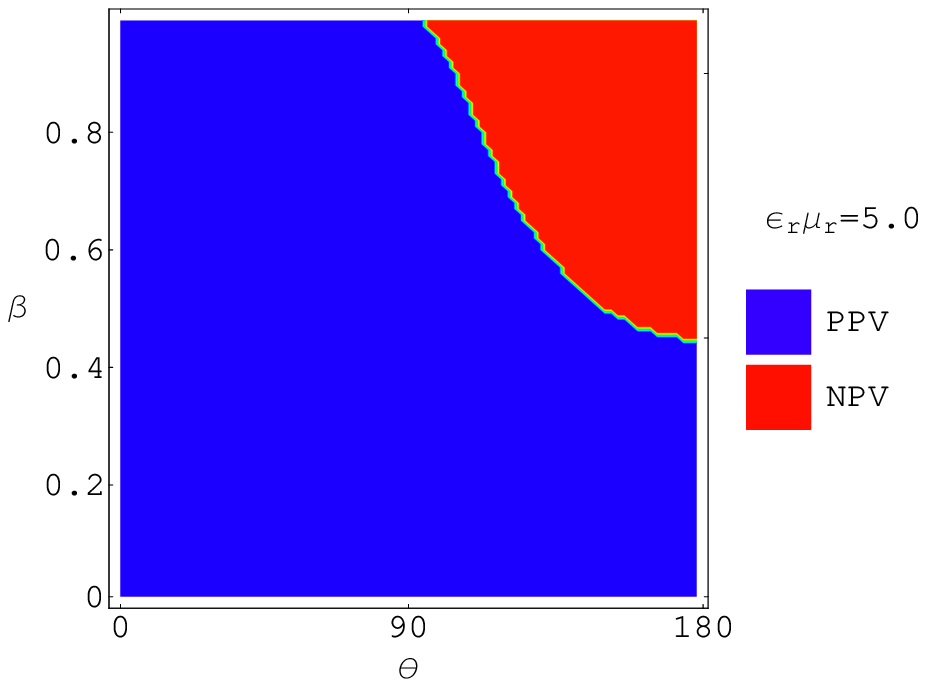,width=2.1in}
\epsfig{file=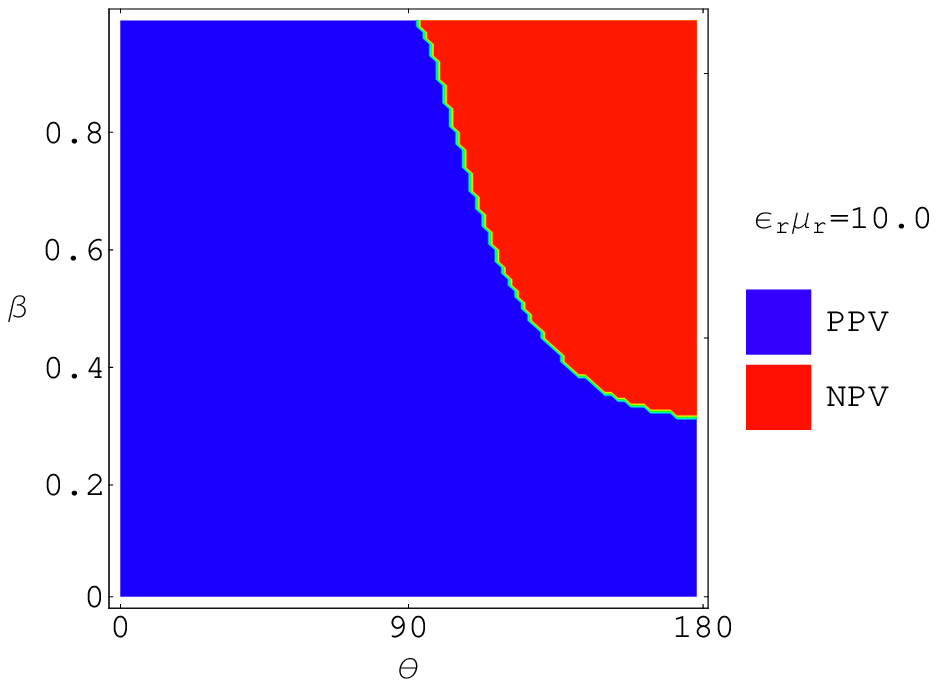,width=2.1in} \caption{continued}
\end{figure}

\end{document}